# STATUS REPORT ON THE LOW-ENERGY DEMONSTRATION ACCELERATOR (LEDA)*

H. Vernon Smith, Jr. and J. D. Schneider,
Los Alamos National Laboratory, Los Alamos, NM 87545, USA

*Abstract*

The 75-keV injector and 6.7-MeV RFQ that comprise the first portion of the cw, 100-mA proton linac for the accelerator production of tritium (APT) project have been built and operated. The LEDA RFQ has been extensively tested for pulsed and cw output-beam currents ≤100 mA. Up to 2.2 MW of cw rf power from the 350-MHz rf system is coupled into the RFQ, including 670 kW for the cw proton beam. The emittance for a 93-mA pulsed RFQ output beam, as determined from quadrupole-magnet-scan measurements, is $\varepsilon_x \times \varepsilon_y = 0.25 \times 0.31$ ($\pi$ mm mrad)$^2$ [rms normalized]. A follow-on experiment, to intentionally introduce and measure beam halo on the RFQ output beam, is now being installed.

## 1 INTRODUCTION

The LEDA RFQ [1] is a 100% duty factor (cw) linac that delivers >100 mA of $H^+$ beam at 6.7 MeV [2-4]. The 8-m-long, 350-MHz RFQ structure [5] accelerates the dc, 75-keV, 110-mA $H^+$ beam from the LEDA injector [6] with ~94% transmission. The primary objectives of LEDA are to verify the design codes, gain fabrication knowledge, understand beam operation, measure output beam characteristics, learn how to minimize the beam-trip frequency, and improve prediction of costs and operational availability for the full 1000- to 1700-MeV APT accelerator. This paper summarizes the RFQ commissioning results given in [1, 3, 4, and 7-13].

## 2 LEDA CONFIGURATION

The accelerator configuration for beam commissioning of the LEDA RFQ is shown in Fig. 1. Major subsystems are the injector [6], ion source and low-energy beam transport (LEBT); RFQ [1, 4, 5]; high-energy beam transport (HEBT) [14]; and the beamstop [15]. The injector (Fig. 2) matches the 75-keV, 110-mA dc proton beam into the RFQ. Simulations of offline injector measurements [16] indicate the RFQ input beam rms normalized emittance is ≤0.23 π mm mrad [6]. A current modulator feeding the microwave magnetron provides beam pulsing [17] for commissioning and beam-tuning activities. The on-line LEBT diagnostics include a pulsed-current toroid, located directly before the RFQ (AC toroid 3), that is used in determining the RFQ transmission.

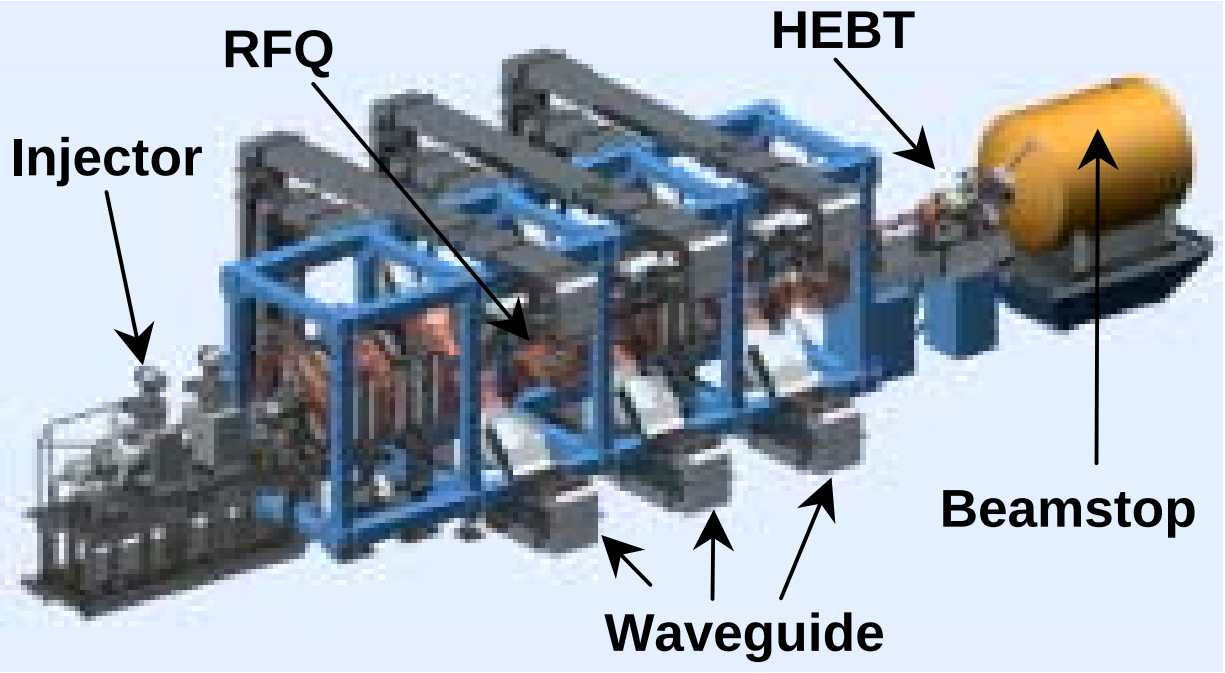

Figure 1. LEDA configuration for RFQ commissioning.

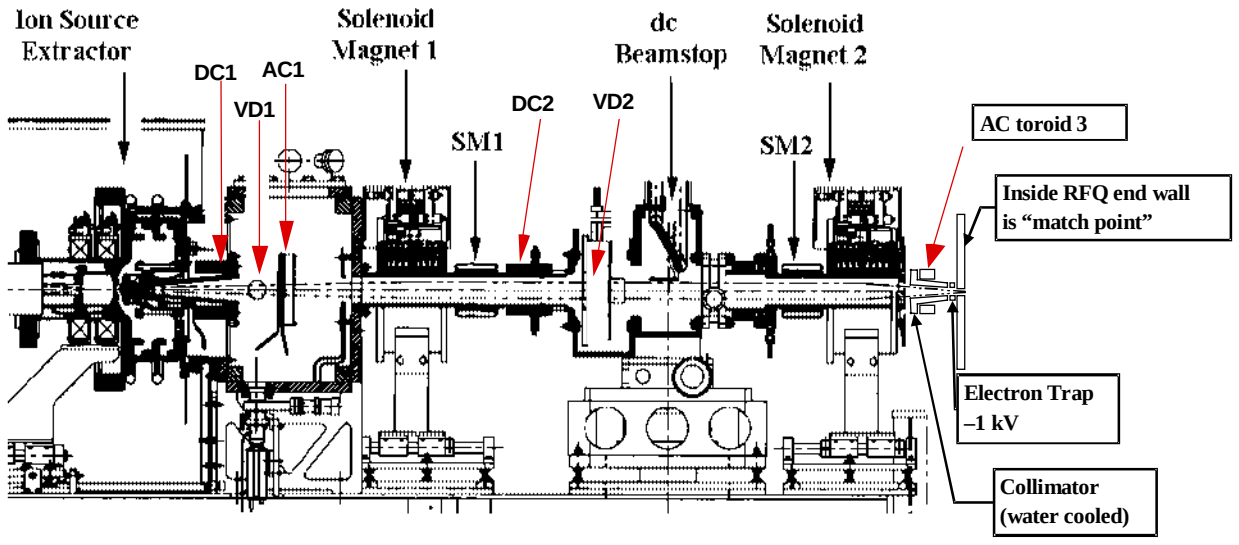

Figure 2. The LEBT beamline with optics and diagnostics.

A complete description of the LEDA RFQ, including the RFQ rf-field tuning procedure, resonance control, and operation with the high-power rf (HPRF) and low-level rf (LLRF) systems, is given in [1, 4, 5, 9, and 11] and the references contained therein. A schematic of the LEDA HEBT showing the location of beamline optics and diagnostics is given in Fig. 3. The function of the LEDA HEBT is to characterize the properties of the 6.7-MeV, 100-mA RFQ output beam and transport the beam with low losses to a water-shielded ogive beamstop [15]. The beamline optics consist of four quadrupole-singlet and two X-Y steering magnets.

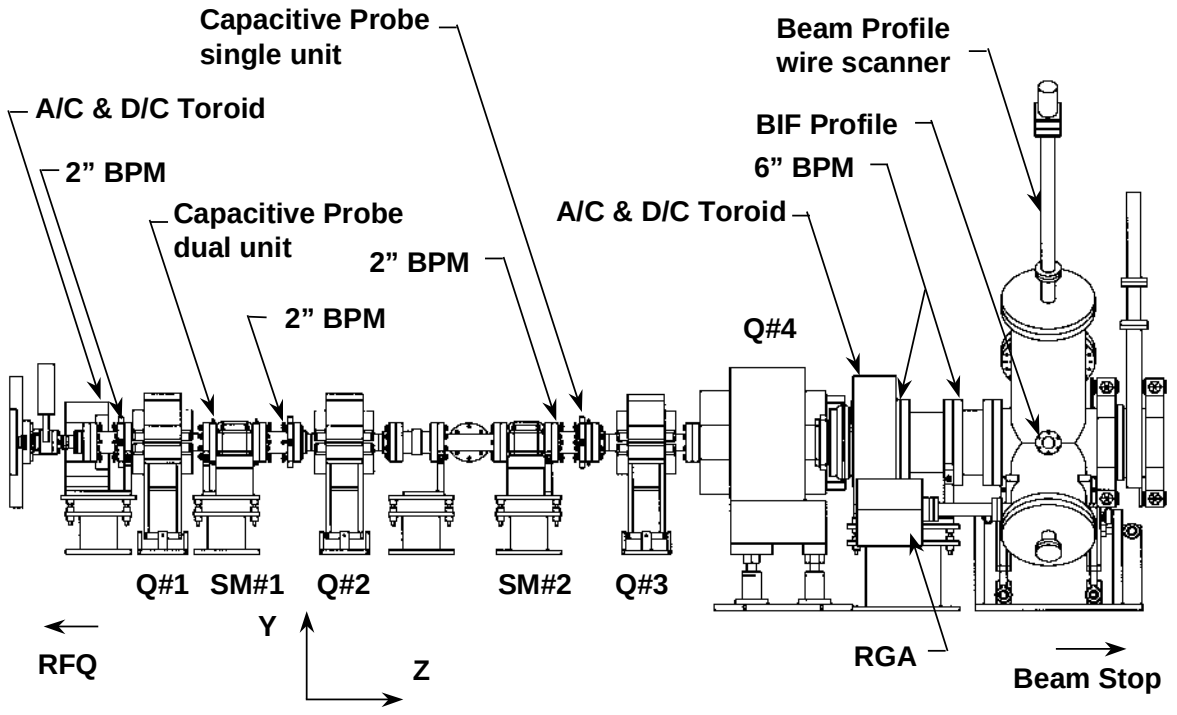

Figure 3. Layout of HEBT beamline optics and diagnostics. Beam direction is from left to right.

* Work supported by the US Department of Energy.

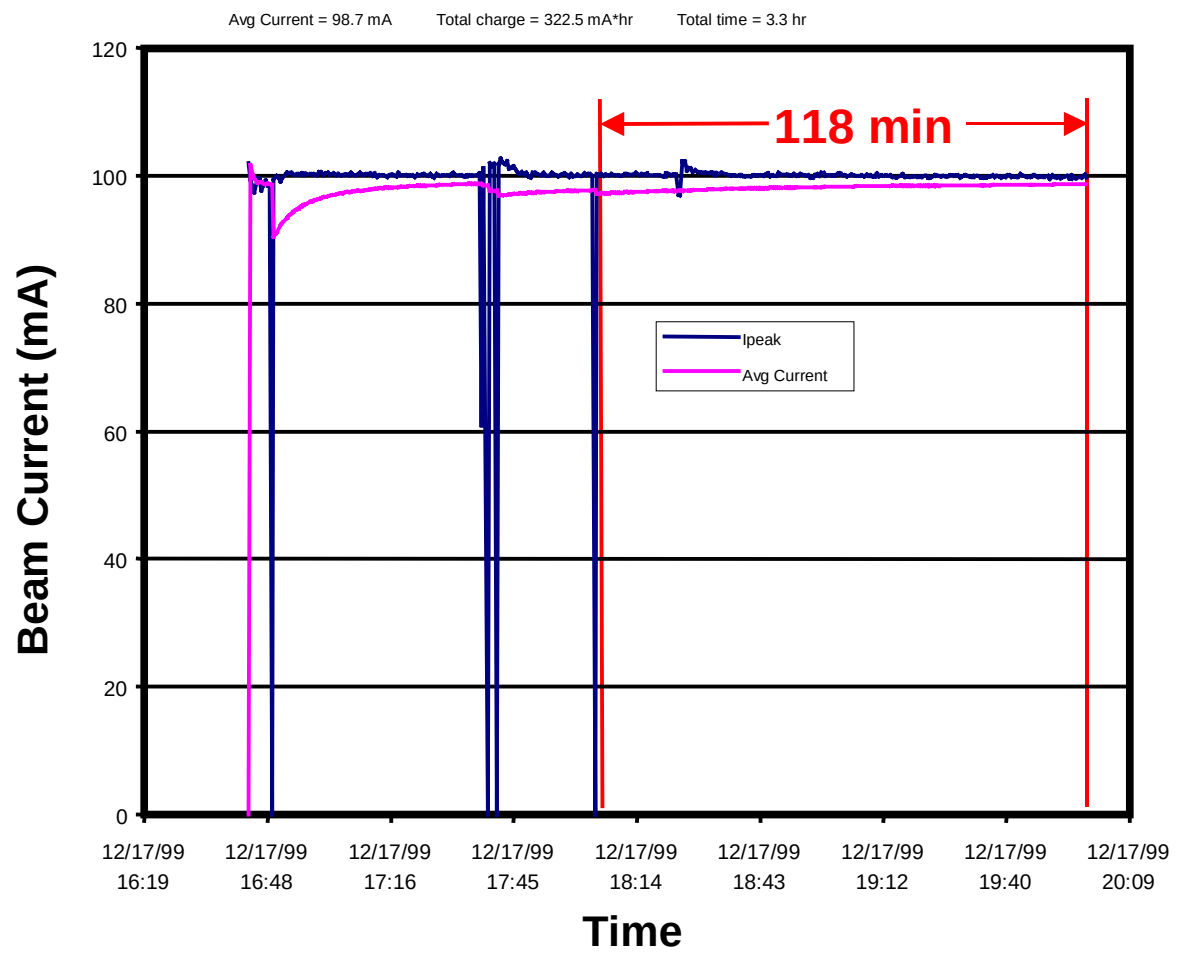

Figure 4. Archived RFQ output beam current (30-s intervals) for a 3.3 hr period on Dec. 17, 1999. Any beam interruptions during the last 118 min are <30-s long.

The HEBT beam diagnostics [18] allow pulsed-beam-current, dc-beam-current, and bunched-beam-current as well as transverse centroid, longitudinal centroid (i.e., beam energy from time-of-flight and beam phase), and transverse beam profile (wire scanner and beam-induced fluorescence) measurements.

## 3 BEAM COMMISSIONING RESULTS AND DISCUSSION

We have accumulated 21 hr of LEDA RFQ operation with ≥99 mA of cw output beam current and >110 hr with ≥90 mA of cw output beam current [11] since modifying the injector and increasing the RFQ rf fields to 5-10% above the design values as described in [1], [4], and [11]. For one run of 118 min (Fig. 4), most of the beam interruptions were 1-6 s in duration (Fig. 5). Recovery from these interruptions, most of them arising from short-duration injector and/or rf-system sparks, was automatic (no operator intervention).

We find that during pulsed beam operation for RFQ rf-field levels at the design value, for pulse lengths >200 μs, and for RFQ output beam currents >90 mA, the RFQ transmission drops abruptly about 100 μs into the beam pulse [11]. The transmission then remains constant at the lower value for the duration of the pulse, including long pulses. The RFQ output beam current for a 300-μs-long beam pulse is shown in Fig. 6. The current abruptly drops by ~10% about 125 μs into the pulse. Figure 7 shows the measured values for the total beam transmission at the start and end of a 500-μs, 2-Hz, 90-mA beam pulse. At the end of the pulse the total transmission deviates from the PARMTEQM prediction for 108-mA output beam current over the field-level range 88-98% of the design (Fig.7). The total transmission at the start of the pulse follows the PARMTEQM prediction for the range 0.91-1.1 of the design rf-field level. For output beam currents >90 mA, e.g. 100 mA, the RFQ transmission over the whole pulse is increased to the design value by increasing the rf-field level to 105-110% of the design field. Both the rf-power system and the RFQ-cooling system allow this increase – the only drawback is that the RFQ requires 10-20% more input power.

The LEDA RFQ output beam emittance is determined [12,13] from quadrupole-magnet scan measurements [12]. For a 93-mA pulsed beam, three x quad scans are given in Fig. 8 and three y quad scans in Fig. 9: also shown are the quad-scan simulations obtained using the particle-optics code LINAC. The LINAC curves in Figs. 8 and 9 are determined by starting with the RFQ output beam Twiss parameters calculated using PARMTEQM, then adjusting these Twiss parameters to give the "fits" to the data shown in Figs. 8 and 9 [12]. The resulting RFQ output beam Twiss parameters are $\alpha_x$ = 1.8, $\alpha_y$ = -2.5, $\beta_x$ = 36 cm, $\beta_y$ = 89 cm, $\varepsilon_x$ = 0.25 $\pi$ mm mrad, and $\varepsilon_y$ = 0.31 $\pi$ mm mrad (rms normalized) [12]. The beam-optics code IMPACT, which, like LINAC, includes non-linear space-charge effects, is used to calculate the beam profiles for each of the quadrupole magnet settings used in the quad scans [13]. LINAC has 2-D (r-z) space charge; IMPACT, 3-D. The Twiss parameters given above are used in these calculations. In Fig. 10 samples for 3 points in the x quad scan are given as a) - c) [Fig. 8] and for 3 points in the y quad scan as d) - f) [Fig. 9]. A computer program that adjusts the Twiss parameters to obtain the best global fit to the measured beam profiles is being considered.

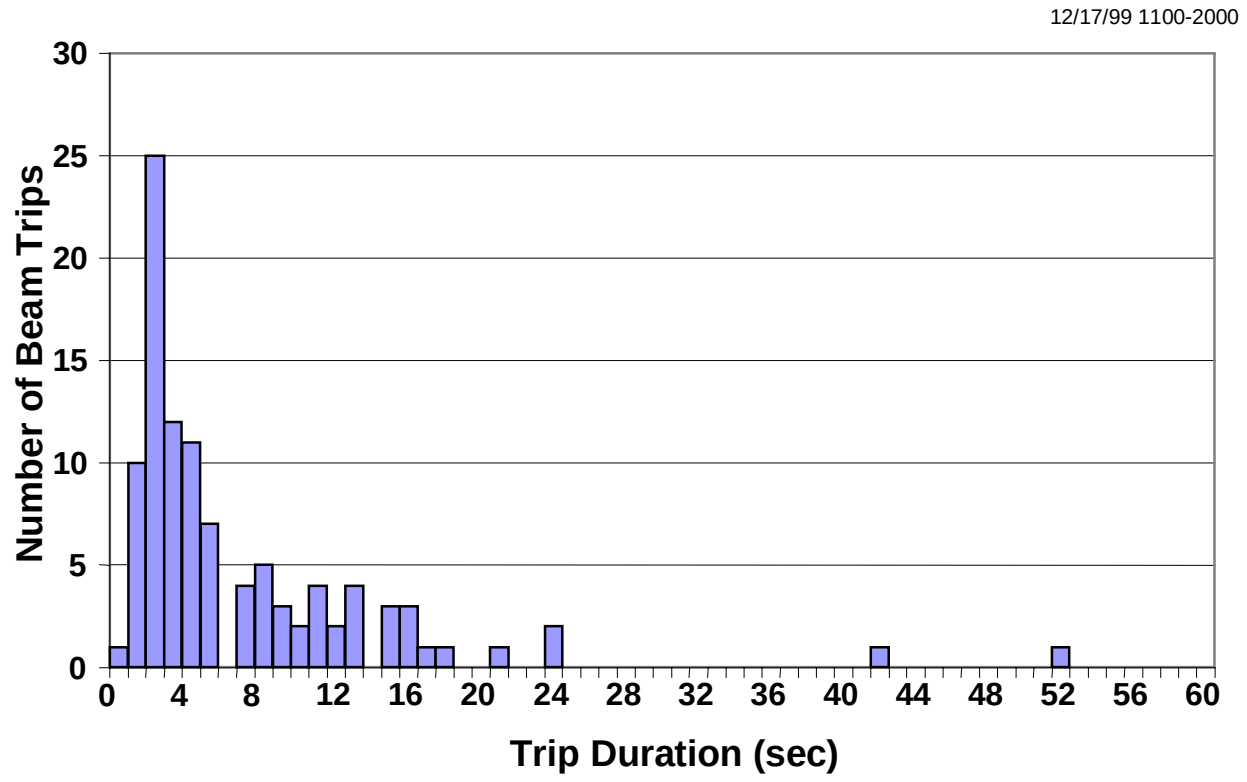

Figure 5. The number of beam trips vs. trip duration (data archived in 1 s intervals) for a 9-hr time period that includes the 3.3-hr interval displayed in Fig. 4.

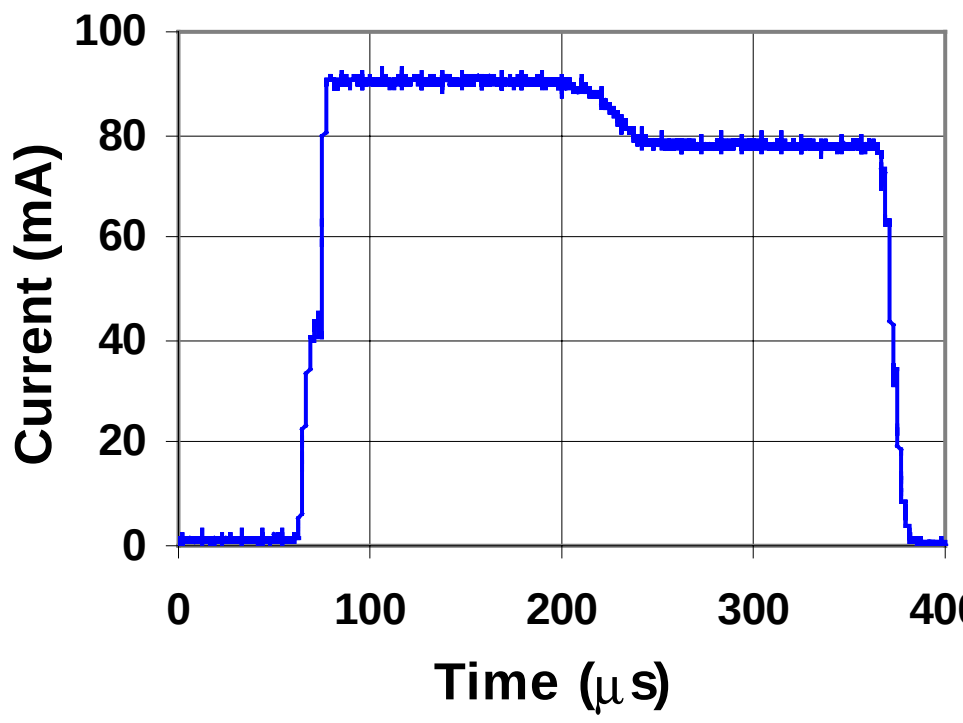

Figure 6. RFQ output beam current vs. time into a 300-μs-long pulse for the design RFQ rf-field level.

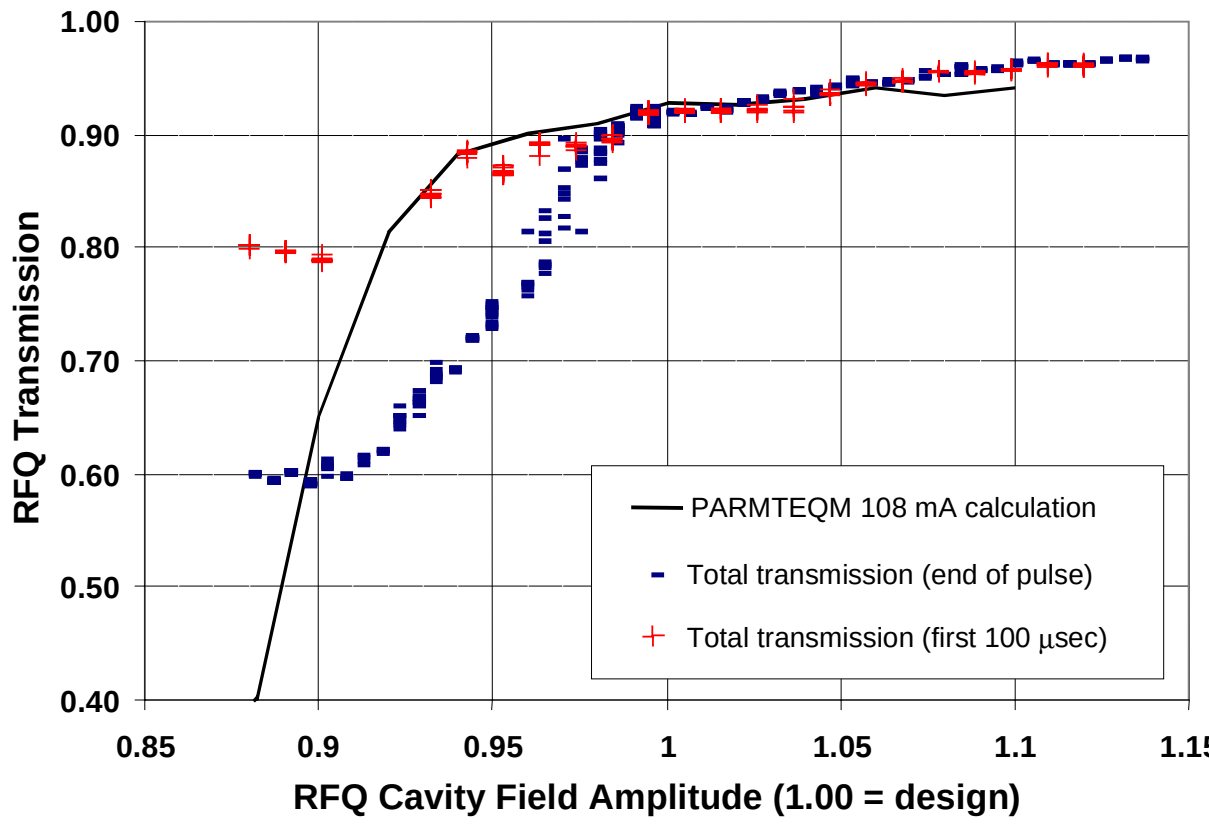

Figure 7. RFQ total beam transmission vs. rf cavity field level at the start (crosses) and at the end (dashes) of a 500-μs-long, 90-mA beam pulse.

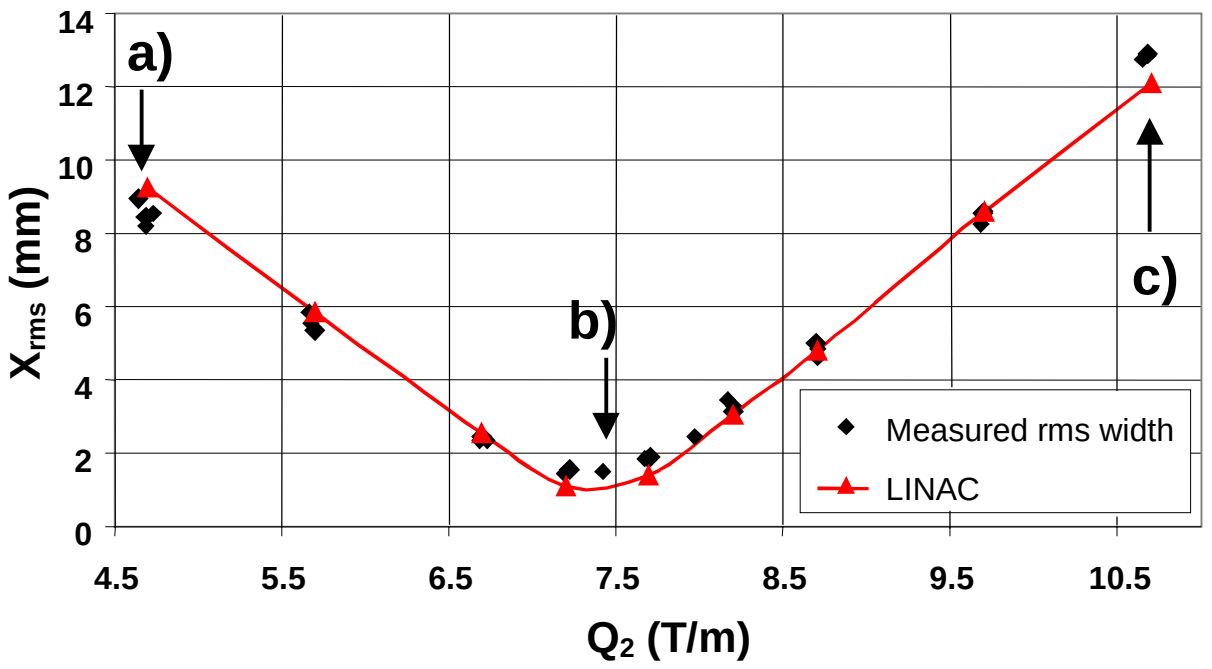

Figure 8. 93-mA x-scan data (diamonds) taken on three different days. The LINAC calculation (triangles, line) has Twiss parameters as described in the text.

## 4 SUMMARY

The LEDA RFQ has operated with ≥99-mA cw output beam for 21 hr cumulative: it has operated >110 hr cumulative with ≥90-mA cw output beam. The RFQ output beam emittance for a 93-mA pulsed beam, determined from quadrupole-magnet-scan measurements, is $\varepsilon_x \times \varepsilon_y = 0.25 \times 0.31\ (\pi\ \text{mm mrad})^2$ [rms normalized]. We are now preparing to intentionally introduce and measure the beam halo in a 52-magnet FODO lattice [19, 20].

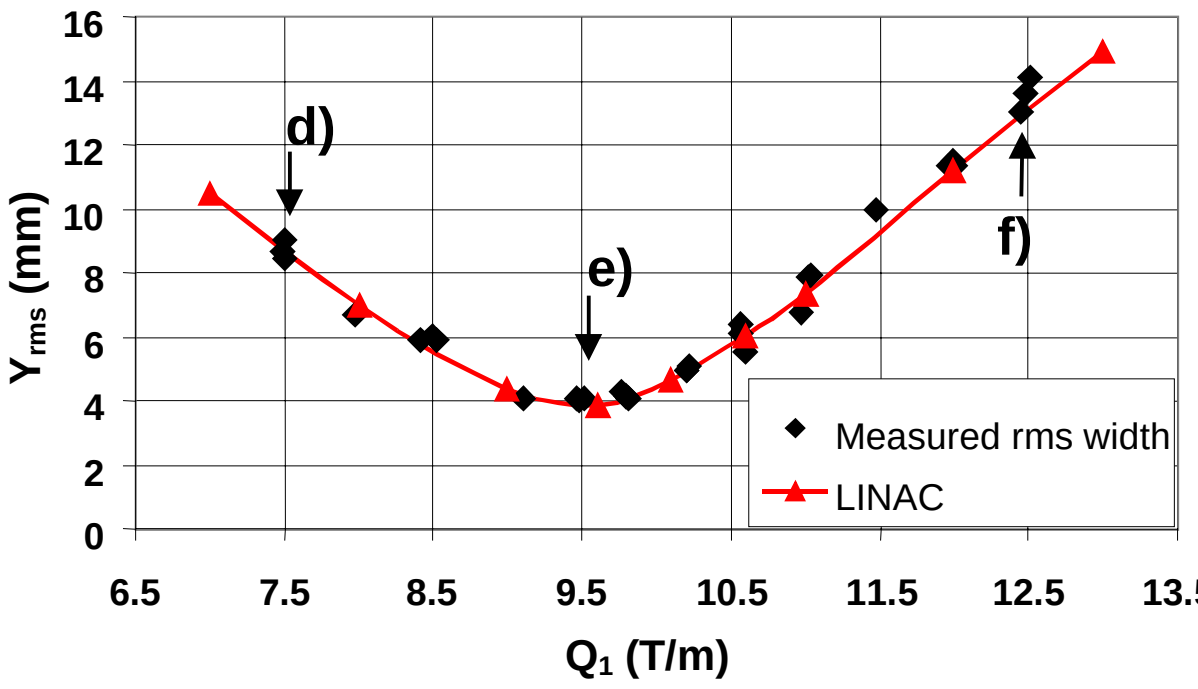

Figure 9. Same as Fig 8, except a y scan.

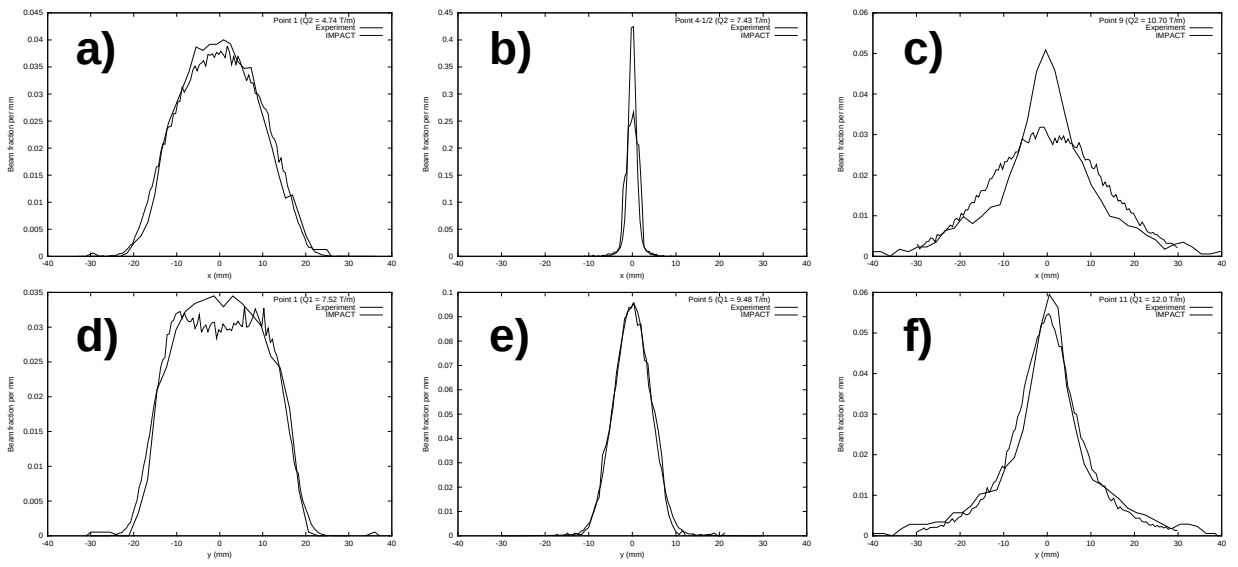

Figure 10. IMPACT calculation (dashed) of the measured (solid lines) 93-mA x- (top) and y-scan (bottom) profiles using the Twiss parameters in the text [12,13].